# AN SDN APPROACH FOR AN ENERGY-EFFICIENT HETEROGENEOUS COMMUNICATION NETWORK IN DISASTER SCENARIOS

Toan Nguyen-Duc and Eiji Kamioka

Graduate School of Engineering and Science, Shibaura Institute of Technology

*ABSTRACT*

*Wireless access technologies have been extensively developed aiming to give users the ability to connect to their expected networks anytime, anywhere. This leads to an increment of the number of wireless interfaces integrated into a single mobile device, hence, it allows the device to be able to connect to multiple access networks. However, in some specific cases such as natural disasters, having an uncorrupted and timely information exchanging means is critical for affected victims to survive or to connect to the outside world. This is because the essential network infrastructures in these cases could be destroyed causing a large number of systems to stop working. In that cases, the victims need a heterogeneous communications network in which they can communicate, without a doubt, by using different wireless access technologies, i.e., Bluetooth or Wi-Fi. The network must also be able to smoothly change the access technologies, or called a vertical handover, to ensure QoS for ongoing applications. In addition, the network must have a mechanism to save energy. For these reasons, an SDN approach, which has been proposed in a previous work, is considered. The performance of the system has been validated by a set of experiments in a real testbed. The obtained results show that the proposed vertical handover can save at least 24.42 per cent of the energy consumed by the wireless communication. The handover delay with different UDP traffic is less than 150ms. Moreover, the network allows a device using Bluetooth to talk with another one using Wi-Fi over a heterogeneous connection where the end-to-end jitter is mainly below 20ms and the packet loss rate is as small as 0.2 per cent.*

## 1. INTRODUCTION

Communication during and right after a disaster plays an important role of the response and recovery as it can connect victims in the stricken area with first responders, support systems, or other family members. Just for example, in catastrophic disasters like the East Japan Earthquake [1], huge areas were damaged, causing crucial network infrastructures to be destroyed. As a result, many systems including the transportation, the cellular services and the electrical power generation were unable to work. Thereby, it is essential to deploy a communication network, which does not rely on any static infrastructured networks, for providing timely information and services. To this end, several researchers have chosen Wi-Fi as a communication means in disaster situations [2, 3, 4]. In [2], the authors have confirmed that the communication network can be formed without relying on infrastructured networks by using software access point (soft-AP). Similarly, in [3] the authors utilized Wi-Fi Direct, which is a new function of Wi-Fi, to connect a mobile device to other devices. Besides, the device can connect to the others in an ad-hoc manner and communicate in an opportunistic way [4]. These studies have shown the feasibility of Wi-Fi during disaster situations in terms of rapid and easy deployment. However, the energy efficiency of the proposed communication network has not been well investigated.

Wi-Fi can operate being independent of any network infrastructures and is easy to use. However, Wi-Fi is an energy consuming protocol, when it is in active state, it consumes at least 15-20% of the total energy capacity of a laptop [5]. It also consumes energy even in sleep state [6]. Besides, disasters are accidental events, hence, the mobile devices would not have been fully charged up in general. It is likely that the battery level of mobile devices is very low, i.e., 20%. In this case, conventional devices, which are communicating using Wi-Fi, will have a very limited operating lifetime. Therefore, it is necessary to have another energy-efficient communication means that allows low battery devices to communicate with the others for a longer period. For this purpose, Bluetooth appears as a high-potential candidate since it is optimized for low power consumption and is designed to enable easy pairing between any two mobile devices. However, the communication rage of Bluetooth is very short, i.e., 10 meters, thus, the Bluetooth connection is easy to be lost, called a failure, due to the mobility of mobile devices. When a failure is about to occur, to maintain the ongoing session between the two mobile devices, they need to switch the communication means from Bluetooth to a larger coverage wireless network such as Wi-Fi. To switch a communication means from one access technology to another while maintaining all the ongoing sessions, a vertical handover is necessary.

To address all the above issues is a non-trivial challenge since it requires a flexible seamless vertical handover to be performed automatically in the environment where there are no infrastructured networks. Moreover, the handover process needs to take into account the operating lifetime of each mobile device. Vertical handover techniques [7, 8, 9, and 10] can be performed on the network layer or higher layers in the OSI model [11]. However, the handover delay is larger than 1000ms [7, 12]. Such a long delay is not recommended for real-time IP-based services, i.e., VoIP application. To reduce the handover delay, several proposals have adopted the method in which all the wireless interfaces on the device operate simultaneously at all times [13, 14]. In [14], the proposal focused only on navigating traffic on all activated, the energy efficiency has not yet been considered. The work in [13] targeted on reducing energy consumption. However, it requires an external element that plays a significant role in the vertical handover process. The external Wi-Fi AP in [13] has been modified so that it can provide Wi-Fi and Bluetooth connections at the same time, allowing the system to switch the network traffic between two interfaces. Likewise, the global controller in [15] has been used to obtain the whole network topology. After having the global view of the network, it can provide the source and destination addresses to the devices for the handover process. The external element in [16], which is called a decision engine, acts as a Media Independent Handover (MIH) server and has the duty of collecting and delivering network information to mobile devices which are requesting to perform a handover. Since the external element can be destroyed or lose the connection to the network in disaster scenarios, the handover processes, which rely on the external element, could not be executed.

To perform a vertical handover automatically under required QoS parameters without any external network devices, our distributed SDN system [17] seems to be a promising solution. In the proposed system, an extended SDN controller has been developed and embedded in each device to collect the network state and to feed it to the main SDN controller. This information helps the main SDN controller to make a right decision in managing the network, i.e., to perform a vertical handover without any support from outside. As an improvement of our previous work, in this paper, a newly redesigned system is proposed to support victims in disaster situations by adding two new criteria counting energy efficiency and robust heterogeneous communication. In terms of energy efficiency, a bidirectional vertical handover algorithm is proposed to reduce the energy consumed by the wireless communication. In terms of robust heterogeneous communication, the system allows two mobile devices to communicate without concerning which access technology they are using. The performance of the proposed system is evaluated by a set of experiments in a real testbed in order to show the validity to real-time IP-based services referring to ITU-T Y.1541 [18]. Additionally, whenever a failure occurs, the extended controller automatically performs a vertical handover to maintain all the ongoing sessions. The captured results show that the handover delay meets the requirement of VoIP applications as recommended by ITU-T G.114 [19].

The rest of the paper is organized as follows: In Section 2, the requirement for the proposed system is present. In Section 3, the new design of an SDN-based mobile device as well as the detail of each operation mode will be explained. Section 4 will describe the evaluation on the performance of the system. Finally, the paper will be concluded in Section 5.

## 2. REQUIREMENTS FOR THE PROPOSED SYSTEM IN DISASTER SITUATIONS

### 2.1. Assumptions in A Disaster Environment

In this work, disaster situations are considered; where public infrastructured networks such as cellular network and WiMAX network cannot work. However, one or more local infrastructured network elements like APs could practically be alive. Regardless of whether or not these APs can provide the Internet access, they could at least bring local connections to nearby devices. In case of no alive APs, any mobile node can become a soft-AP [2] to establish communication channels for the others. In addition, victims' devices such as smartphones and tablets can join a Bluetooth network since Bluetooth is available on those devices as a standard function. These tentative networks should effectively be used for local communications.

When evacuating, the victims tend to potentially go together to form a group and to assist others [20, 21, 22 and 23]. In this case, they often need to use a real-time application such as VoIP by their mobile devices as a critical communication means rather than passing messages for rapidly making a right decisions [24]. Therefore, the voice stream must be sustained until exiting the affected area. In such a VoIP network, a call server is commonly required. However, the call server is just to process the call setup, and thus it does not switch the voice stream. Once the call server finishes setting up a call, it generally becomes dormant during the conversation. Therefore, the mobile devices should know the other's IP address to have a peer-to-peer call without working with a call server.

### 2.2. Energy-Efficient Wireless Communications

The US Department of Homeland Security (DHS) has a program named SAFECOM [25] in which technical requirements for communication services in disasters have been introduced. The key aspects in the requirements include Quality of Service (QoS), energy consumption, robustness and reliability. Therefore, the proposed communication system needs to meet these requirements.

- Energy efficient communication:

  Wi-Fi is designed for high-speed wireless data transmissions. Although the standard has already introduced a power-saving mode (PSM) for saving energy, the power consumption in Wi-Fi is mostly linear with the obtained throughput [26]. Besides, the WLAN interface consumes energy even if it is in the sleep state [6]. In contrast, the devices can operate for years if they use Bluetooth technology [27]. Therefore, turning off Wi-Fi and switching the on-going sessions to Bluetooth are reasonable for an energy efficient communication. To realize this real-time session switching, a fast vertical handover algorithm is necessary.

- Robust heterogeneous communication:

  The proposed system must be able to maintain all ongoing sessions while changing the access technologies. For instance, if two devices connecting each other via Bluetooth are going to be disconnected due to their distance, the system must be able to switch the ongoing traffic to Wi-Fi with low packet loss rates. The proposed system also should be reliable adapting to the change of network environment including the device mobility, resulting in providing a robust heterogeneous communication. In the heterogeneous communication environment, all the devices are able to connect to the others without nervousness about what technology they are using. The connection in the network must meet the required QoS parameters for real-time services such as VoIP.

## 3. SDN CONTROLLER DESIGN FOR COMMUNICATION MANAGEMENT IN DISASTER SCENARIOS

### 3.1. Traditional SDN Controller

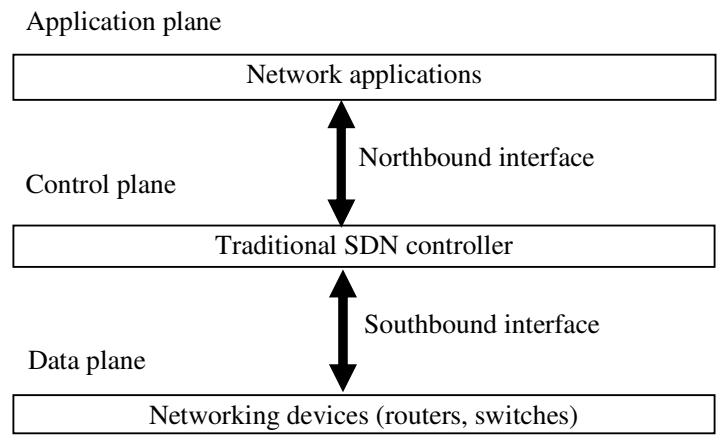

Fig. 1. Traditional SDN architecture

Figure 1 shows a traditional SDN architecture [28] in which the control and data planes are decoupled. In the data plane, networking devices, i.e., switches, work as simple packet forwarding elements, thus sometimes the data plan is called forwarding plane. The control function of network devices is moved to an external element, called SDN controller. The traditional SDN controller (traditional-SDNC) has two key functions. First, it is responsible for controlling or forwarding packets in the data plane. The packets are forwarded in a flow-based mechanism. In the mechanism, all the packets that match the criteria defined in one flow rule will be processed in the same way, forming a flow. To check the matching, when packets reach the networking devices, their header field values are taken out and are compared to the stored flow rules. The flow rules are typically defined by the traditional-SDNC and then are sent to the networking devices via the soundbound interface (SBI). Second, the traditional-SDNC collects status information from networking devices to offer a global view of the network to applications in the application plane. The traditional-SDNC communicates with the application via the northbound interface (NBI). The interface also allows the traditional-SDNC to listen to requirements from each network application (nwApp). After received the requirements, the traditional-SDNC interprets them into the instructions, which will be used to control the data plane.

**3.2. The Need of an Extended SDN Controller**

As previously mentioned, a traditional-SDNC typically controls the packets that go through the networking devices, i.e., OpenFlow switches. The traditional-SDNC is commonly located separately from the switches and has a global network view including the network topology, link status, etc. Therefore, to control the packets, the controller only needs to talk with the switches. However, when a traditional-SDNC and an OpenFlow switch are embedded in one node of the network as shown in Fig. 2, the controller may not be able to have the global network view due to the network disconnection and it is unable to answer any requests from the switch. Therefore, it is necessary to enrich the features of the local traditional SDN controller (local-SDNC) so that it can control over the OpenFlow switch.

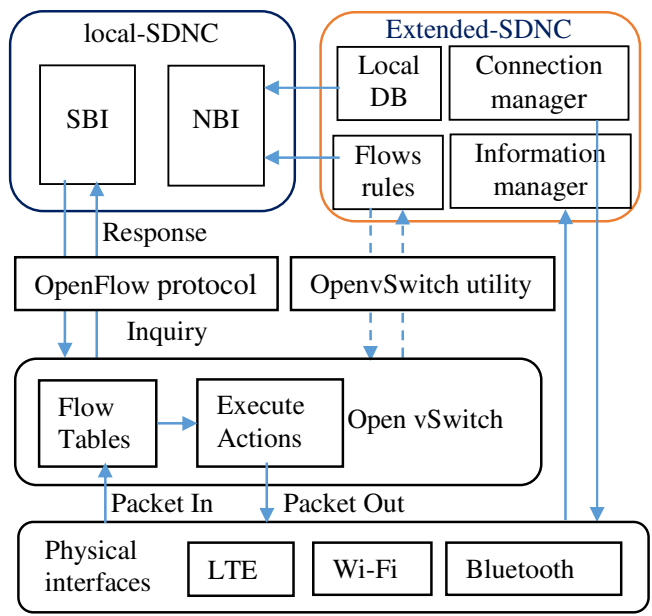

Fig. 2. Control logic of an SDN-based mobile node

To control the switch, a local-SDNC needs at least information of the network such as the network topology. To learn the network topology, the network information including IP and MAC addresses must be exchanged among the devices. Therefore, the first feature that needs to be enhanced for the local-SDNC is a device-to-device (D2D) communication. Since the local-SDNC's design is to talk only with the networking devices, i.e., OpenFlow switches, the messages exchanged in a D2D communication must go through the switches. Note that the traditional-SDNC mainly uses OpenFlow protocol to talk with the switches. In the OpenFlow specification [28], there is no option to allow a traditional-SDNC to send any messages through switches to another traditional-SDNC. However, it is possible that a local-SDNC instructs the switches to forward and receive a message, which is not generated by the local-SDNC, to another local-SDNC. Since nwApp in SDN architecture is to steer the switches through a traditional-SDNC. Therefore, it is recommended to use a non-SDN software, or an extended SDN controller (extended-SDNC), to play the role of exchanging D2D's messages.

In case a link failure occurs, the proposed system must perform a vertical handover to maintain the end-to-end connection. When performing the vertical handover, all the ongoing traffic are switched from one physical interface to another. This action requires the reconfiguration of the network to detach an access technology and attach to the other one. Since the traditional-SDNC and nwApps are targeted at controlling the switches, the extended-SDNC should be involved in managing the physical network configuration. Besides, the change of physical interface will change the IP and MAC addresses in the packet's header, hence, the system must ensure that any running application is not interrupted during and after the vertical handover process. Not to let the applications to detect the change of physical interface, the system needs to provide a virtual interface with unchanged IP and MAC addresses for them. To this end, the packets, which are exchanged between two end devices, must have those values in their header rewritten. SDN technology already offers the packet's header rewriting feature. Note that to send or receive data, the devices must exchange up–to–date ARP packets, this feature again requires the extended-SDNC.

Since a local-SDNC is a computer software, it can be hung or killed while the other computer software keeps running. When the local-SDNC dies, nwApps will lose their interpreter to translate their abstract requirements into the commands, which will be executed on the switches. Different from nwApps, the extended-SDNC is designed as a separated program from traditional-SDNC, thus, it can be in charge of controlling the switch while waiting for the local-SDNC to be recovered.

### 3.3. The Extended SDN Controller Design

The extended-SDNC is designed to cover all requirements for disaster situations, described in the previous section. Different from the former design [17], the new design has three new components named "Flow rules", "Connection manager" and "Information manager" as shown in Fig. 2. With these components, the extended-SDNC has three new features. The first new feature, which is implemented in the "Information manager" component, is to collect all the necessary information. The information includes not only the network configuration information such as MAC and IP addresses, access point properties, but also the network traffic information like throughput, round-trip time (RTT). The "Information manager" component gathers information from the local side as well as exchanges the collected information with other devices. When the SDN-based mobile device is not connected to any network, the component collects all local information such as network properties, MAC address and IP subnet of all interfaces. The information is then stored in a local database called "Local DB". Once the device is connected to the network and starts to communicate with any device, the "Information manager" component on each device will exchange and update their local database information. The "Information manager" component only performs this process one time between two end devices at the beginning of their communication.

The second new feature coming with the extended-SDNC is that it plays a role of a temporary controller when the local-SDNC has problems in controlling the switch. For instance, when the local-SDNC is frozen or being killed, the Open vSwitch [29] will have no controller to ask. In this situation, the extended-SDNC needs to be in charge of controlling the switch at least until the local-SDNC comes back. In fact, the extended-SDNC itself can form Open Flow rules based on the information in the "Local DB". The rules are formed and then sent to the switch by the "Flow rules" component. This component is able to talk with the switch simply because it utilizes a built-in utility of the Open vSwitch called ovs-vsctl.

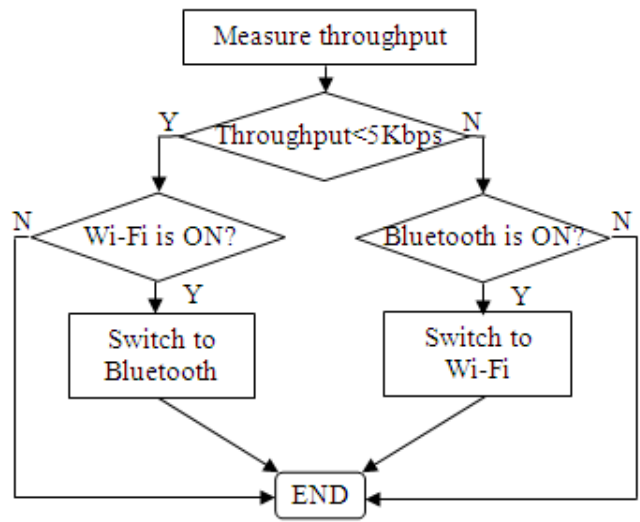

Fig. 3. A vertical handover triggering algorithm

Here, the third new feature of the extended-SDNC appears. The new feature is implemented in the "Connection manager" component and is specially designed for the disaster scenarios in which the mobile devices need to save energy as much as possible almost all the time. The component is to decide if the communication interface switching from the current access technology to the other is needed or not. The collected information can also be used to connect wireless mobile devices, which are using different access technologies. In other words, the feature enables a heterogeneous communication means in which mobile devices do not need to care about the access technology that they are using.

By having the above three new features, the extended-SDNC allows the SDN-based mobile devices to work in two modes, namely, energy-efficient communication mode and robust heterogeneous communication mode.

### 3.3.1. Energy-Efficient Communication Mode

This mode is for conserving battery life of the device. Besides the basic methods such as dimming the brightness of the display screen and disabling certain applications, turning off or disabling unconnected wireless interfaces will allow the mobile device to use much lower power during periods of inactivity. The key reason for focusing on this is that wireless technologies constantly seek for a good connection by scanning, hence, they consume the energy continuously. Note that the feature of switching between the activated wireless interface and the backup one without disturbing the running application has been introduced in [17]. Different from the proposed system in [17] which mainly focused on the handover execution procedure, here, when and how the handover is triggered is discussed more concretely.

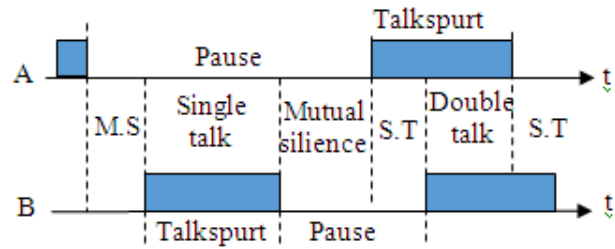

Fig. 4. Conversational speech model in ITU-T P.59

The algorithm for triggering the energy-efficient vertical handover is given in Fig. 3. Firstly, the network throughput is measured by "Connection manager" component. The component acquires the network throughput by utilizing the Python psutil library [30]. In this work, if the data rate for communication is lower than 5Kbps, it is considered as no traffic. If there is no traffic for a certain period in second (hereafter this is called ThresholdWB), this component will check which wireless interface is currently activated. If the activated interface is Wi-Fi, the "Connection manager" component will try to switch the connection to Bluetooth for saving energy. In case Bluetooth interface is used, it will be maintained. Note that if at least one of the communicating two devices is out of the Bluetooth communication coverage, the Wi-Fi connection will be maintained. The ThresholdWB is selected based on ITU-T P.59 [31] in which VoIP traffic is half duplex because Wi-Fi is also half duplex. In WLANs, radio channels are using 2.4 GHz or 5 GHz. However, APs commonly share one radio channel for both uplink and downlink because they apply an access control method called CSMA/CA (Carrier sense multiple access with collision avoidance). As a result, Wi-Fi enabled devices communicate in a half-duplex manner where they either transmit or receive at any given time to avoid collisions. As shown in Fig. 4, the listener typically responses after a pause period. Thereby, it is assumed that if there is no response in two pause periods, it is possible to switch the connection to Bluetooth to save energy. Note that the Bluetooth connection still allows the users to continue the conversation, but the voice quality may be poor. Since the average of a pause period is 1.587s as given in [30], the ThresholdWB is selected as 3s. Similarly, if the data rate is greater than 5Kbps in ThresholdBW seconds, the "Connection manager" component will also check which access technology is being used. If it is Bluetooth, it will be switched to Wi-Fi for better voice quality. The value of ThresholdBW is selected as the period of time to let both sides hear the other's voice. As shown in Fig. 4, ThresholdBW is equal to twice the sum of the silence time (Mutual Silence) and the talking time (Talkspurt). In ITU-T P.59, the average Mutual Silence (M.S) and Talkspurt are 0.508s and 1.004s, respectively, thus, ThresholdBW is determined as 3s. Also, if the switching process fails, the Bluetooth connection is maintained.

International Journal of Wireless & Mobile Networks (IJWMN) Vol. 8, No. 6, December 2016On the next step, the "Connection manager" component prepares for switching the connection. Based on the backup access network, a corresponding association procedure is generated and saved in a shell file waiting for the call from the "Connection manager" component. OpenFlow rules are also defined and described so that they can be sent to the local-SDNC as well as directly to the switch if necessary. The information of MAC and IP addresses for the backup interface are also prepared. For the sake of simplicity, it is assumed that transferring packets, i.e., ARP, ICMP, TCP, UDP and IP, in the network are already known. Traffic classification is out of scope in this work.

```
Get current activated interface
Switch to another interface based on
activated one
  Case "Bluetooth"
    Turn Wi-Fi interface on
    Associate Wi-Fi interface to an access
point
    If association process finished
      Perform synchronization process
    End If
  End case
  Case "Wi-Fi"
    Turn Bluetooth interface on
    Join a Bluetooth piconet
    If pairing process finished
      Perform synchronization process
    End If
  End case
Open a thread to execute network
configuration
Open a thread to install OpenFlow rules
Update the local database
```

Fig. 5. Handover execution

```
Timer is ON
If received the SYN message from the
other device
// which already finished its association
process
  Notify him with a SYN message
  Do {wait for partner response
Else
  Notify him with a SYN message
  Sleep in RTT/2
End if
```

Fig. 6. Synchronization process

To execute a handover between two wireless access technologies, the "Connection manager" component on each device makes the mobile device abandon the current connection and get attached to the other network, which is selected in the previous step. To ensure no packet is lost in this process, several protocols have been standardized [7, 8, 9, 10]. The more complex in this procedure is, the more network overhead is generated, and the longer the handover delay is. Therefore, in this work, a simple socket message exchange procedure, which allows both sides to understand the other's status, is used. Since the procedure is simple, it reduces the handover delay. The detail of this procedure is illustrated in Fig. 5. Firstly, the "Connection manager" component on each device checks which interface is being used and which one is in the backup state. The "Connection manager" component then wakes the backup interface up and associates it with its corresponding network. After finishing the association process, the device must enter the synchronizing step, which is shown in Fig. 6, to make sure the other side has also finished this association process and to allow both sides to switch to the other access network at the same time. Before the association process begins, a timer is set in case there is an interference between Bluetooth and Wi-Fi. When the interference occurs, the exchanged packets may get lost or spend a long time to reach the destination. In that case, when the timer counts down to zero, the association process is over. Otherwise, in a normal association process, the device first checks if it receives any message from the other one to verify whether that device finishes the association process on the backup interface yet. If there is no message, meaning that the other side has not finished associating the interface, the device sends a SYN message to him informing that this side has finished the association process. After that, the device waits for the other side device's response. Note that the traffic still goes through the active interface during this time. In case the other side device has already finished the association process, he will reply with a SYN message as the confirmation and then wait for a half of RTT to let the message reach the destination. After both sides have received the SYN message, each one performs network configuration and installs new OpenFlow rules. Note that these processes need to perform at the same time to minimize the handover delay, thus, multithreading or multiprocessing technique is used. As a result, the handover delay will be the largest in the period from when configuring the network to when installing the rules. When the network configuration process finishes, the extended-SDNC must send the ARP packets to the other side advertising the new interface with new MAC and IP addresses. After the ARP packets are exchanged, two devices can continue to communicate via the new interface.

### 3.3.2. Robust heterogeneous communication mode

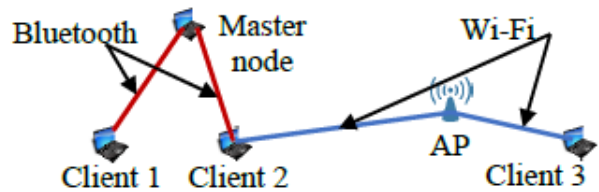

Fig. 7. Schematic of the sustainable heterogeneous communication network

Table 1: Average energy consumption of wireless network interfaces

| Interface | Wake-up (mJ) | Sleep (mW) | Active (mW) | Off (mW) |
|---|---|---|---|---|
| Wi-Fi | 536.77 | 495.05 | 660.09 | 213.75 |
| Bluetooth | 417.98 | 79.24 | 104.21 | 38.20 |

Although tethering, which is supported in both Wi-Fi and Bluetooth, allows sharing the Wi-Fi or Internet connection of the mobile device with other devices, it cannot maintain the connection when the device changes the access technology. To this end, this mode is designed as an additional function to the energy-efficient communication mode to enable two SDN-based mobile devices to communicate without concerning which technology they are using. The target of this mode is to allow low battery devices, which need Bluetooth technology to prolong the operating lifetime, to communicate with the others. One case study for this mode is depicted in Fig. 7. The figure shows that on the left side, there are three nodes named Client1, Client2 and Master node which are communicating in a Bluetooth piconet. On the right side, Client2 connects to the Client3 via the Wi-Fi AP. To allow Client1 to talk with Client3 without changing the access network, Client2 volunteers as a repeater node. On Client2, both Bluetooth and Wi-Fi interfaces are turned on and all incoming packets from Client1 and Client3 are passed via Wi-Fi and Bluetooth interface, respectively. Note that while Client2 acts as a repeater, it is still able to communicate with other nodes in the piconet via Bluetooth interface as well as with Client3 via Wi-Fi. This is feasible because all physical interfaces of Client2 are under control of the Open vSwitch. The switch navigates traffic by asking a local-SDNC. The local-SDNC refers to the collected network information in the "Local DB" provided by the extended-SDNC to correctly answer the switch. Once received the response from the local-SDNC, the switch saves the rule as an entry in its flow table. Thereafter, any packets that meet the stored flow rules are forwarded through the switch without referring to the local-SDNC. Moreover, the Open vSwitch has ability to rewrite the packets' header, hence, the source and destination MAC and IP addresses of the incoming packets from an interface can be rewritten and sent to the other interface. Note that to enable the packets to reach their destination, ARP packets have already been handled by the extended-SDNC.

## 4. PERFORMANCE EVALUATION

In this section, two modes of an SDN-based mobile device are evaluated. Experiments were conducted in a real testbed in which several Linux computers were connected via Wi-Fi or Bluetooth. Each computer was equipped with Core 2 Duo @2.26 GHz processor and 2GB RAM. On each computer, two USB-based wireless adapters including a wireless LAN adapter and a Bluetooth 4.0 one have been installed for accessing Wi-Fi and Bluetooth networks, respectively.

### 4.1. Evaluation of Energy-Efficient Communication Mode

This mode is specially designed for energy saving, thus, the evaluation focuses on estimating how much energy can be saved.

#### 4.1.1. Measurement of Energy Consumption

In this section, the energy consumption in different states of wireless network interfaces, i.e., Wi-Fi and Bluetooth is presented. The states in the experiment include "wake up", sleep, active and off. The "wake up" state starts when the interface is activated and finishes when the interface has completed the association process. The sleep state in this work is defined as the state where the mobile device has connected to a wireless network, but no applications on the device are using the wireless connection. In the event of active, the interface can transmit or receive the network traffic which has the rate higher than 5Kbps. When the interface is off, it is still attached to the mobile device, but it is disabled.

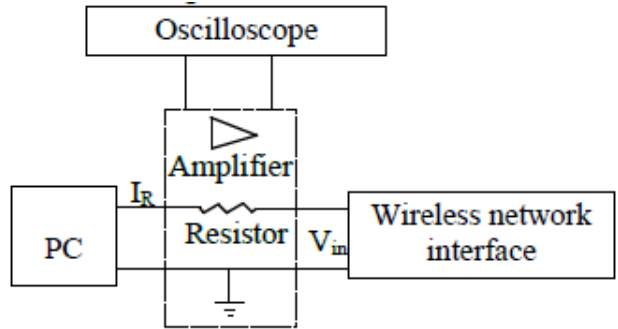

Fig. 8. Measurement setup to determine energy consumption of wireless network interfaces

The experiment setup to determine the power consumed in each event is illustrated in Fig. 8. In the figure, a wireless network interface is connected to a Linux computer, which has been equipped with the proposed system, via a 0.5 ohm shunt resistor. The obtained current $I_R(t)$ needs to be amplified before being sent to the oscilloscope since the value of the current can be very small in case of Bluetooth. The voltage $V_{In}(t)$ dropped on the network interface is measured directly. The instantaneous energy consumption $P(t)$ of the wireless interface is computed based on the Ohm's law: $P(t)=I_R(t)V_{In}(t)$. After that, the average energy consumption is calculated, similarly to the technique used in [6], within a period of time that approximately wraps a state. For example, the average time for Bluetooth interface to be woken up and be associated with the master node is $T_{Wakeup}$, thus, the average energy consumed in this state is the average value of $P(t|0 \rightarrow T_{Wakeup})$. The measured results of the average energy consumption of each wireless interface in different events are shown in Table 1. As seen from the table, for completing the "wake up" process, Bluetooth consumes 417.98 millijoule while Wi-Fi consumes more energy, which is 1.2 times of Bluetooth. The comparison unit is millijoule since Wi-Fi and Bluetooth complete the "wake-up" process in different $T_{Wakeup}$ time. The average value of measured $T_{Wakeup}$ of Wi-Fi and Bluetooth are 1.4s and 3.03s, respectively. In the sleep and active states, Wi-Fi consumes energy more than 6 times of Bluetooth. Besides, although the interface has been disabled in the off state, it still consumes energy.

### 4.1.2. Evaluation of Energy Saving

This evaluation is to show how the proposed system saves the energy consumed by the wireless communication. In the system, the energy-efficient vertical handover algorithm allows the wireless connection to be switched from Wi-Fi to Bluetooth whenever there is no traffic. When there is no traffic, the power-saving mode (PSM) will turn Wi-Fi to sleep state to consume $P_{wi-Fi-sleep}$=495.05mW. Although the Bluetooth interface is disabled, it still consumes $P_{Blue-off}$=38.20mW. The average energy consumed by the wireless communication is $P_{wireless} = P_{wi-Fi-sleep} + P_{Blue-off} =$ 533.25mW. This means the wireless communication consumes 533.25mJ every second. Assumed that at $t_0$=1s the system starts the vertical handover process. The energy consumed by the wireless communication at $t=t_0$ is 533.25mJ as shown in Fig. 9. Before the device can use Bluetooth connection, Bluetooth interface must be woken up and be associated with the master node. This process lasts $t_{Blue-wakeup}$=3.03s on average and consumes $P_{Blue-wakeup}$=417.98/3.03=137.95mW. In addition, during this time, Wi-Fi is still in sleep mode and consumes $P_{wi-Fi-sleep}$ (495.05mW). The total energy consumed at $t_1=t_{Blue-wakeup}$ is $P_{wireless}(t|t_0 \rightarrow t_1) = (137.95+495.05)\times 3.03 = 1917.99$mJ. The consumed energy in this period is higher than the energy consumed by the sleeping Wi-Fi interface and Bluetooth interface, which is in the off state, $P_{\text{Wi-Fi-only}}(t|t_0 \rightarrow t_1)=(495.05+38.20) \times 3.03=1615.75.12$ mJ.

After Bluetooth is woken up, Wi-Fi is disabled, then the total energy consumption will be $P_{\text{wireless}}(t|t_1 \rightarrow t_1+t_{\text{Blue-sleep}})=(79.24+213.75)t_{\text{Blue-sleep}}$ (mJ) if Bluetooth interface sleeps in $t_{\text{Blue-sleep}}$ seconds. In the worst case, there is some traffic in the network right after the handover process has finished. In that case, $t_{\text{Blue-sleep}} = 0$.

When there is traffic in the network, Bluetooth interface switches to active state and consumes the energy $P_{\text{Blue-active}}(t|t_1 \rightarrow t_1+t_{\text{Blue-active}})=104.21 t_{\text{Blue-active}}$ (mJ). Due to the proposed algorithm, the time needed to decide for switching back to Wi-Fi is 3s, hence, $t_{\text{Blue-active}}=t2=3$ seconds. During this period, the disabled Wi-Fi interface consumes $P_{\text{Wi-Fi-off}}=213.75$ mW. The total energy consumed by the wireless communication is $P_{\text{wireless}}(t|t_1 \rightarrow t_2)= (104.21+213.75) \times 3=953.88$ mJ. If the system does not use the proposed algorithm, Wi-Fi connection will be used all the time. Then, the energy consumed when Wi-Fi is active is $P_{\text{Wi-Fi-only}}(t|t_1 \rightarrow t_2)= (660.09+38.2) \times 3=2094.87$ mJ.

To switch the connection back to Wi-Fi, Bluetooth interface is disabled, Wi-Fi interface is woken up and is associated with the AP. The average time for Wi-Fi to complete this process is $t_{\text{Wi-Fi-wakeup}}=t_3=1.4$ seconds. The system consumes $P_{\text{Wi-Fi-wakeup}}=536.77/1.4=383.41$ mW to wake the Wi-Fi interface up and associate it with an AP. The total amount of consumed energy during this time is $P_{\text{wireless}}(t|t_2 \rightarrow t_3)=(383.41+38.2) \times 1.4=590.25$ mJ. The energy consumed by using only Wi-Fi is much higher $P_{\text{Wi-Fi-only}}(t|t_2 \rightarrow t_3)=(660.09+38.2) \times 1.4=977.60$ mJ. The total amount of energy consumed by only Wi-Fi is $P_{\text{Wi-Fi-only}}(t|t_0 \rightarrow t_3)=4593.54$ mJ while the energy consumption by using the proposed algorithm in the worst case is $P_{\text{wireless}}(t|t_0 \rightarrow t_3)=3471.66$ mJ. This means, the proposed system has saved 24.42% of the total energy consumption by the wireless communication. The Fig. 10 shows the amount of energy saved when the $t_{\text{Blue-sleep}}$ increases. For example, the system saves 44.7% energy consumption when the $t_{\text{Blue-sleep}}=600$ s.

The obtained results have confirmed that in any cases, the proposed system can save the energy consumed by the wireless network interfaces.

## 4.2. Evaluation of Robust Heterogeneous Communication Mode

### 4.2.1. Evaluation of the System's Robustness

In order to evaluate the system's robustness, the network performance has been observed when a vertical handover occur. The evaluation metrics is the handover delay, which is defined in Section 3 as the largest in the time period from when configuring the network to when installing the Open Flow rules.

The testbed in this evaluation consists of three SDN-based mobile devices as illustrated in Fig. 11. The figure shows that the devices are communicating in a Bluetooth piconet for saving energy. It is assumed that all the devices accept the participations from other devices. The devices are also able to join the 802.11g network in this testbed by utilizing the built-in Wi-Fi interface. The 802.11g protocol was selected for the test as the Wi-Fi connection simply gives a faster transmission rate and a larger communication coverage. In the evaluation, Iperf [32] is used to generate an UDP flow toward the direction from Client1 to Client2. The execution times on both Client1 and Client2 were captured by utilizing a Python module named "time" and the results are given in Fig. 12. As shown in the figure, for each process, the average and 95% confidence interval of the execution times were measured. The measured results with two nodes are similar. In case the connection is going to be switched from Bluetooth to Wi-Fi, both $t_{WOF}$ and $t_{WNW}$ for Wi-Fi network are around 80ms and the difference between them is small. In the invert direction, $t_{BNW}$ varies from 80ms to 150ms and $t_{BOF}$ is around one sixth of $t_{BNW}$. The larger the difference between $t_{BOF}$ and $t_{BNW}$ is, the larger the number of packet loss is. The packets are lost because they are navigated to another interface due to the newly installed OpenFlow rules. However, the interface has not been ready yet. Ideally, the configuration and the installation should be processed at the same time, and thus the packet loss does not occur.

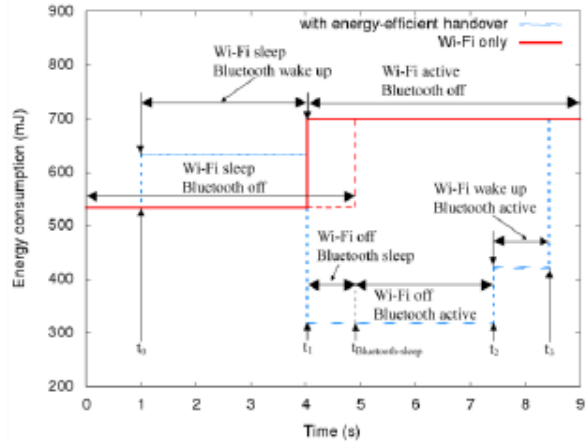

Fig. 9. Energy saved when performing a vertical handover

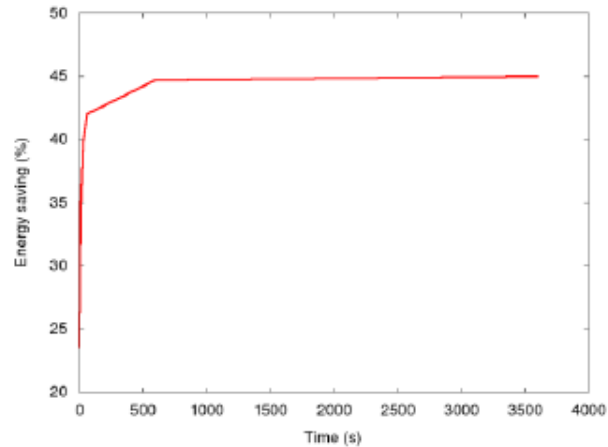

Fig. 10. Energy saved by using proposed system

The experiment results have confirmed the flexibility of the proposed system in the bidirectional vertical handover between Bluetooth and Wi-Fi with different UDP traffic load. In addition, the handover disruption time, which causes the handover delay, was measured in the experiment. The result shows that the handover delay was still less than 150ms, which is acceptable for VoIP applications as recommended by ITU-T Recommendation G.114.

#### 4.2.2. Evaluation of the Heterogeneous Communication Network

To evaluate the quality of the heterogeneous communication network in supporting real-time services, i.e., VoIP data, the QoS parameters including packet loss rate and jitter have been measured. These parameters were selected because packet loss in VoIP will typically reduce the quality of the voice communication. The variation in the delay of received packets, or jitter, is also a reason of discarded packets if it is so large.

Figure 7 shows the real testbed in this evaluation. In the figure, on the right side of topology, the Client2 and Clinet3 are attached to the AP, through the 802.11g Wi-Fi interface. On the left side, the Client2 is also a member of a Bluetooth piconet which has a Master node and another client named Client1. The Client2`s duty is to pass the traffic between Client1 and Client3.

In the evaluation, UDP traffic from Client3 to Client1 was generated using Iperf. The UDP protocol was selected because the VoIP data is placed in the packets of Real Time Protocol (RTP) and RTP normally uses UDP as the default transmission protocol. The traffic throughput was varied from 100Kbps to 500 Kbps. In this experiment, the QoS parameters including jitter and packet loss rate were observed since they were selected as the evaluation metrics. Each experiment was repeated 30 times and the results are presented in Fig. 13 and Fig. 14.

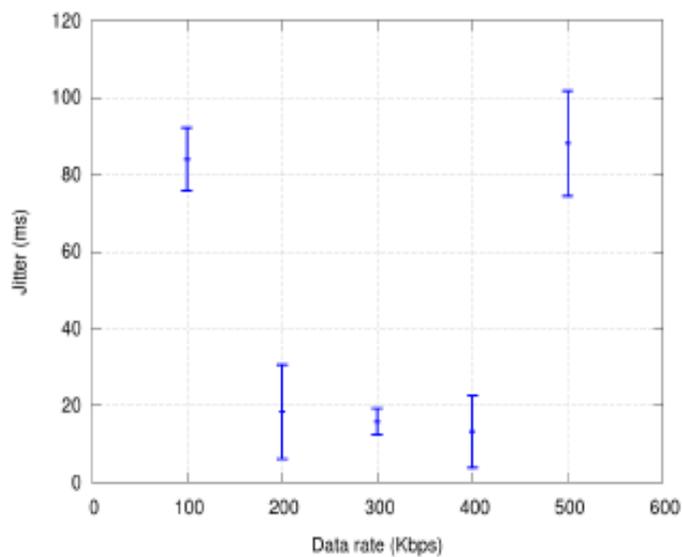

Fig. 13. End-to-end jitter of the decent heterogeneous communication network

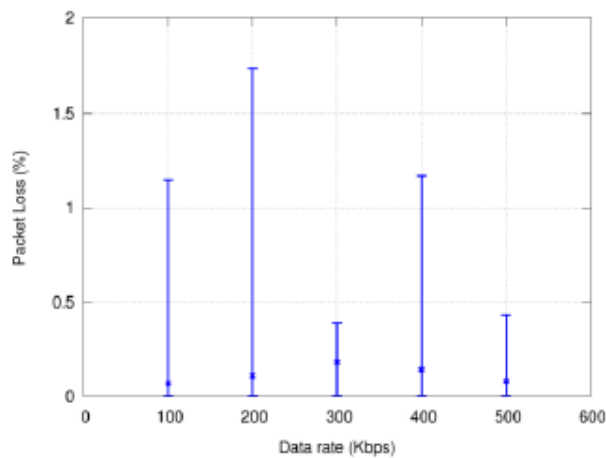

Fig. 14. Packet loss rate of the decent heterogeneous communication network

Figure 13 shows that when the throughput was between 200 Kbps and 400 Kbps, the average end-to-end jitter was less than 20ms. When throughput was 500Kbps or very small, i.e., 100Kbps, the jitter was around 85ms. This is because there was congestion when the traffic throughput was high. In contrast, when the traffic throughput was low, the number of sending packets is small, hence, if a few packets are delayed in reaching the destination, the average delay still increases. Besides, when the data rate is 300Kbps the jitter is more stable and less than 20ms.

Figure 14 shows the end-to-end packet loss rate of the heterogeneous communication network. As shown in the figure, the minimum packet loss rate is zero with any data rate. Although the rate is fluctuated, it is under 0.2% on average. When the data rate is 300 Kbps, the loss rate is more stable and the maximum packet loss rate is 0.39%.

The obtained results have indicated that the system can provide a heterogeneous communication network with the packet loss rate of less than 0.2% and the jitter of less than 20ms when the data rates is varied from from 200 Kbps to 400 kbps. The results also pointed out that the proposed network meets the requirement of real-time IP-based services as recommended by ITU-T Y.1541.

## 5. CONCLUSIONS

In this paper, an SDN-based approach for providing an energy-efficient heterogeneous communication network for victims in disaster scenarios has been introduced. Also, the performance of the system including two operation modes, that is to say, energy saving and robust heterogeneous communication modes, has been evaluated on four major criteria: energy saving, pack loss rate, jitter and handover delay. The system succeeded in dynamically performing handovers between Bluetooth and Wi-Fi in any direction when the traffic was travelling across the system. The handover delay was as small as 85ms when switching the connection from Bluetooth to Wi-Fi. In the invert direction, the delay was larger, however, it was still less than 150ms which is the maximum value of one-way latency recommended by ITU-T G.114. Thereby, the proposed system can save at least 24.42 per cent the energy consumed by the wireless communication. Moreover, the system can allow a mobile device using Bluetooth to talk with another one using Wi-Fi over a heterogeneous connection. The measured end-to-end jitter and packet loss rate show that the proposed network is acceptable for real-time IP-based services referring to ITU-T Y.1541.

International Journal of Wireless & Mobile Networks (IJWMN) Vol. 8, No. 6, December 2016

## Authors

**Toan Nguyen-Duc** is currently pursuing his PhD in Shibaura Institute of Technology, Japan. He received the B.S. and M.S. degrees in electronic and telecommunication from Hanoi University of Science and Technology. Before studying at SIT, he is a researcher at Hanoi University of Science and Technology. His current interests include software defined networking, mobile computing, wireless networks and wireless network communication protocols.

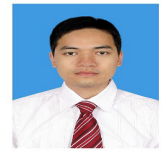

**Eiji Kamioka** is a Professor at Shibaura Institute of Technology. He received his B.S., M.S. and D.S degrees in Physics from Aoyama Gakuin University, Japan. Before joining SIT, he was working for SHARP Communications Laboratory, Institute of Space and Astronautical Science (ISAS) as a JSPS Research Fellow and National Institute of Informatics (NII) as an Assistant Professor. His current research interests encompass mobile multimedia communications and ubiquitous computing.

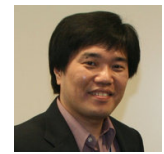